# Prefetching of VoD Programs Based On ART1 Requesting Clustering

P Jayarekha *,  Dr.T R GopalaKrishnan Nair**

* Research Scholar, Dr. MGR   University  Dept. of   ISE, BMSCE, Bangalore .

Member, Multimedia   Research Group, Research Centre, DSI, Bangalore.

** Director, Research and Industry Incubation Centre, DSI, Bangalore.

Jayarekha2001@yahoo.co.in     trgnair@yahoo.com

*Abstract*

*In this paper, we propose a novel approach to group users according to the VoD user request pattern. We cluster the user requests based on ART1 neural network algorithm. The knowledge extracted from the cluster is used to prefetch the multimedia object from each cluster before the user's request. We have developed an algorithm to cluster users according to the user's request  patterns based on ART1 neural network algorithm that offers an unsupervised clustering. This approach adapts to changes in user request patterns over period without losing previous information. Each cluster is represented as prototype vector by generalizing the  most frequently used URLs that are accessed by all the cluster members. The simulation results of our proposed clustering and prefetching algorithm, shows enormous increase in the performance of streaming server. Our algorithm  helps the server's agent to learn user preferences and discover the information about the corresponding  sources  and other similar interested individuals.*

**1 Introduction**

For the past few years, Multimedia applications are growing rapidly and we have witnessed an exponential growth of traffic on the internet [11]. These applications include Video-on demand, Video authoring tools, news broadcasting, Videoconferencing digital libraries and interactive video games. The new challenges that has raised today is concerned with aspects of data storage, management processing, the continuous arrival of requests in multiple, rapid, time-varying and potentially unbounded streams. It is usually not feasible to store the request arrival pattern in a traditional database management system in order to perform delivery operation on the data later on. Instead, the request arrival must generally be processed in an online manner from the cache which also holds the predictive prefetched video streams and guarantee that results can be delivered with a small   start up  delay for the first-time access videos. The VoD streaming server is an important component as it is responsible for retrieving different blocks of different video stream and sends them to different users simultaneously. This is not an easy task due to the real time factor and the large volume of characteristics possessed by the video. Real time characteristic requires that video blocks have to be retrieved from the server's disk within a deadline for continuous delivery to users. Failure to meet the deadline will result in jerkiness during viewing [13].

With the rapid development in VoD streaming services, the knowledge discovery in the multimedia services has become an important research area. These can be classified broadly into two classes multimedia content mining and multimedia usage pattern mining. An important topic in learning user's request pattern is the clustering of multimedia VoD users, i.e, grouping the users into clusters based on their common interest. By analyzing the characteristics of the groups, the streaming server will understand the users better and may provide more suitable, customized services to the users. In this paper, the clustering of the users request access pattern based on their browsing activities is studied. Users with similar browsing activities are clustered or grouped into classes (clusters). A clustering algorithm takes as input a set of input vectors and gives as output a set of clusters and a mapping of each input vector to a cluster. Input vectors which are close to each other according to a specific similarity measure should be mapped to the same cluster[5,8]. Clusters are usually internally represented using prototype vectors which are the vectors indicating a certain similarity between the input  vectors  are mapped to a cluster. Automated knowledge discovery in large multimedia databases is an increasingly important research area.

VoD application   services over a computer network. It provides to watch any video at any time. One of the requirements for VoD system implementation is to have a VoD streaming server that acts as an engine to deliver videos from the server's disk to users. Video blocks should be prefetched intelligently with less latency time from the disk and hence service high number of streams. However, due to real time and large volume characteristics possessed by the video, the designing of video layout is a challenging task. Real time constraints the distribution of blocks on disk and hence decrease the number of streams being delivered to users.

The deficiency possessed by the existing prefetching techniques like window based prefetching and active prefetching methods and cumulative prefetching[14]  is that these schemes only perform prefetching for the currently accessed object and perform prefetching for the currently accessed object and prefetching is only triggered when the client starts to access that object. For the first time accessed objects, its initial portion will not be fetched by current caching and prefetching schemes. So client suffers from start-up  delay for the first-time access.

 In this paper, we consider the problem of periodic clustering and prefetching of first-time access video streams using ART1 neural networks [10]. A target is set as the request arrival pattern, to achieve the results comparative with that of target the videos are prefetched using a ART1 model for which a set of videos to be prefetched are grouped on time-zone.

*Outline of the Paper*: The paper is organized into various sections as follows: section 2 provides some back-ground information, both on VoD streaming model and clustering. Section 3 discusses about the related work in clustering and prefetching. Section 4 presents the methology used in developing the algorithm. Section 5 evaluates our proposed model through analysis and simulation results. Section 6 presents conclusions.

## 2 Background

### 2.1 The VoD Streaming Model

The VoD streaming model has the following characteristics:

- The request arrival pattern is online and they might be undergoing on some specific constraints.

- The order in which the request arrives is not under the control of the system.

- Request arrivals that have been processed are either discarded or serviced. They cannot be re retrieved easily unless being stored in the cache memory, which is just the prefix of the whole video stream.

- The prefetching operation should have lower priority than normal request. So the video steaming server should prefetch only when its workload is not heavy.

- The prefetched objects are required to content with normally fetched objects for cache space.

### 2.2 Proposed architecture

Multimedia streaming servers are designed to provide continuous services to clients on demand. A typical Video-on-demand service allows the remote users to play any video from a large collection of videos stored on one or more servers. The server delivers the videos to the clients, in response to the request.
Multimedia Streaming Servers, specifically customized for HTTP, RTSP based streaming are ideally suited for developing quick I.P. based streaming systems.
Multimedia streaming setup, as shown in Figure 1, includes two types of interactions.
The streaming server processes the real-time multimedia data and sends it to the clients, through the various possible types of devices.
In the server part, the multimedia streaming server accepts multimedia data or input from any of the following sources:

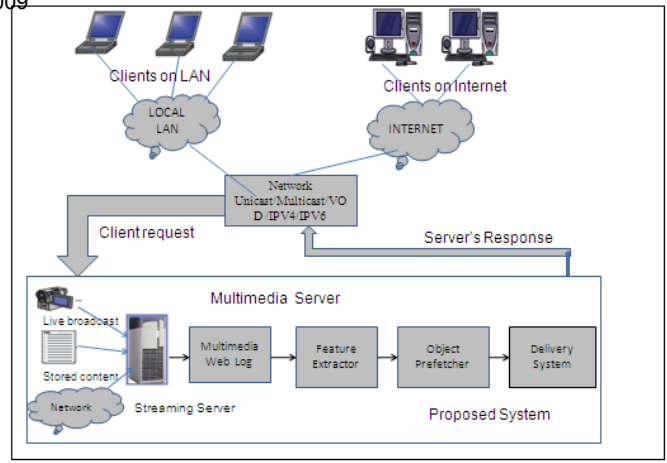

**Fig 1 Multimedia Streaming Server**

**Architecture**

• Live broadcast, such as a digital camera connected to the computer port.

• Data stored in the form of media.

• Data stored on machines in a network.

The system targets Remote Monitoring, live event broadcasting, home/office monitoring, archiving of the video in a centralized server and related applications. Streaming multimedia data is a transaction between the server and client. The client, can be any HTTP client, that accesses the media. The server is an application that provides all the client applications with the multimedia content. Unlike the download-and-play mechanism, the multimedia streaming client starts playing the media packets as soon as they arrive, without holding back to receive the entire file. While this technology reduces the client's storage requirements and startup time for the media to be played, it introduces a strict timing relationship between the server and the client.

### 2.3 Clustering

A cluster is a collection of data objects that are similar to one another within the same cluster and are dissimilar to the objects in other cluster. A cluster of data objects can be treated collectively as one group and so may be considered as a form of data compression. Also, cluster analysis can be used as a form of descriptive statistics, showing whether or not the data consists of a set of distinct subgroups.
The term cluster analysis (first used by Tryon, 1939) encompasses a number of different methods and algorithms for grouping objects of similar kind into respective categories. A general question facing researchers in many areas of inquiry is how to organize observed data into meaningful structures, that is, to develop taxonomies. In other words cluster analysis is an exploratory data analysis tool which aims at sorting different objects into groups in a way that the degree of association between two objects is maximal if they belong to the same group and minimal otherwise. Clustering and unsupervised learning do not relay on predefined classes and class-labeled training examples. For this reason, clustering is a form of learning by observation, rather than learning by examples.

## 2.4 Clustering Video Streams

The clustering of (active) video streams concerns the concept of distance or, alternatively, similarity between streams. It is required to decide how two video streams fall into one cluster. Here, we are interested in the time-dependent evolution of a request generated to a video stream. That is to say, two different request to video streams are considered similar if their evolution over time shows similar characteristics. The customer behavior change over time the accurate predictions are rather difficult. A successful video streaming server is the one which offers customer with large selection of videos. While prefetching the videos from the server's disk and clustering them in one cluster the different factors that are considered are, The time at which the requests are generated. For example children's videos are likely to be popular early in the evening or in weekend mornings, but less popular late at night. The maximum total revenue the service provider can make is limited by the capacity of the server and number of active videos that are currently present in the cache, and hence the videos that are clustered into one should generate not only maximum revenue but also reduce the waiting time. The videos can be categorized into children videos, adult videos, house-wife videos and hot news videos. Thus the video steaming system should adopt rapidly and service the request using predictive prefetch to a widely varying and highly dynamic workload.

## 2.5 Advantages of neural networks and ART1 in clustering and prediction

Neural network is highly tolerance of noise data, as well as their ability to classify pattern on which they have not been trained. They can be used when we may have little knowledge of the relationships between attributes and classes.
Artificial neural networks are taught trained to perform specific functions in contrast to conventional computers that are programmed. Training in neural networks could be supervised or unsupervised. In supervised learning training is achieved using a data set by presenting the data as input-output pair of patterns, and the weights are accordingly adjusted to capture the relationship between the input and response. Unsupervised network are on the other hand are not aware of desired response and therefore learning is based on observation and self-organization.
Adaptive Resonance Theory an unsupervised learning gives solutions for the following problems arising in the design of unsupervised classification systems [12]

- Adaptability : Refers to the capacity of the system to assimilate new data and to identify new clusters (this usually means a variable number of clusters)
- Stability: Refers to the capacity of the system to conserve the clusters structures such that during the adaptation process the system does not radically change its output for a given input data.

Neural networks can be used for prediction with various levels of success. The advantage of this includes automatic learning of dependencies only from measured data without any need to add further information (such as type of dependency like with the regression). The neural network is trained from the historical data with the hope that it will discover hidden dependencies and that it will be able to use them for predicting into future. In other words, neural network is not represented by an explicitly given model.

In the arena of multimedia communication, knowledge of demand and traffic characteristics between geographic area is vital for the planning and engineering of multimedia communications. Demand for any service in a network is a dynamically varying quantity subject to several factors. Prior knowledge of request patterns for various services will allow utilization of storage resources. This prior knowledge is essential for planning of multimedia server architecture for the future in an optimized manner, as well as for allocating available storage in an effective fashion, based on the current demand. In this paper we have proposed a generalized scheme for learning the behavior of video on demand characteristics on a daily and weekly basis using simulation data.

## 3 Related work

We discuss the significant work in the area of clustering and prefetching.

### 3.1 Related work in clustering

Clustering users based on their Web access patterns is an active area of research in Web usage mining. R. Cooley et al. [4] propose a taxonomy of Web Mining and present various research issues. In addition, the research in web mining is centered on extraction and applications of cluster and prefeching. Both of these issues are clearly discussed in [1].It has been proved in this scheme a 97.78% of prediction hierarchy. Clustering of multimedia request access pattern is defined by hierarchical clustering method to cluster the generalized session. G T Raju et al. [5] has proposed a novel approach called Cluster and PreFetch (CPF) for prefetching of web pages based on the Adaptive Resonance Theory (ART) neural network clustering algorithm. Experiments have been conducted and results show that prediction accuracy of our CPF approach is as high as 93 percent.

### 3.2 Related work in prefetching

Prefetching means fetching the multimedia objects before the user request them. There are some existing prefetching techniques, but they possess some deficiency. The client will suffer from start-up delay for the first-time accesses since, prefetching action is only triggered when a client starts to access that object. However, an inefficient prefetching technique causes wastage of network resources by increasing the web traffic over the network.[6] J Yuan et al has proposed scheme, in which proxy servers aggressively prefetch media objects before they are requested. They make use of servers' knowledge about access patterns to ensure the accuracy of prefetching, and have tried to minimize the prefetched data size by prefetching only the initial segments of media objects. [7] KJ Nesbit et al has proposed an prefetching algorithm which is based on a global history buffer that holds the most recent miss addresses in FIFO order.

## 4. Methology

### 4.1 Preprocessing the web logs

The multimedia Web log file of a web server contains the raw data of the form <client_Id, User_Id, timestamp, requested_video,HTTP replay code,bytes sent>. The format of the log file is not suitable for directly applying ART1 algorithm on them. Therefore, the log data needs to be transformed into a <client_Id,date, requested_video> format. We have selected 50 clients requesting for 200 different videos.

### 4.2 Extraction of feature vectors

For clustering, we need to extract the binary pattern vector P that represents the accessed requested_video's URL by a client and is an instance of the base vector $B =\{URL_1,URL_2.....URL_n\}$. The pattern vector maps the access frequency of each base vector element $URL_i$ to binary values. It is of the form $P =\{P_1,P_2,....P_n\}$ where each $P_i$ is either 0 or 1. $P_i$ is 1 if $URL_i$ is requested by a client 2 or more times, it is 0 otherwise.

| 0 | 1 | 1 | 0 | 1 | 1 | 1 | 0 | 1 | 1 |
|---|---|---|---|---|---|---|---|---|---|

**Fig 2 Sample Pattern Vector**

Fig 2 is a sample of pattern vector generated during a session. Each pattern vector has a binary bit pattern of length 200. For each session we input 50 such pattern to an ART1, since we have 50 clients.

### 4.3 Session Identification

A user requesting for a video may visit a Web site from time to time and spend arbitrary amount of time between consecutive visits. To deal with the unpredictable nature of generation of request arrival pattern, a concept called session is introduced. We cluster the request pattern in each session. The data collected during each session is used as historical data for clustering and prefetching purpose. Subsequent requests generated during a session is added as long as the elapse of time between two consecutive requests does not exceed a pre_defined parameter *maximum_idle_time*. Otherwise, the current session is closed and a new session is created.

### 4.4 Clustering users Adaptive Resonance Theory

Figure 3 depicts the ART1 architecture and its design parameters $b_{ij}$ (bottom-up weights) is the weight matrix from the interface layer to the cluster layer, $t_{ij}$ (top-down weights) is the weight matrix from the cluster layer to the interface layer, N(number of input patterns or interface units), M (maximum allowed number of clusters) and ρ(vigilance parameter). The vigilance parameter ρ and the maximum number of clusters M has big impact on the performance[8]:

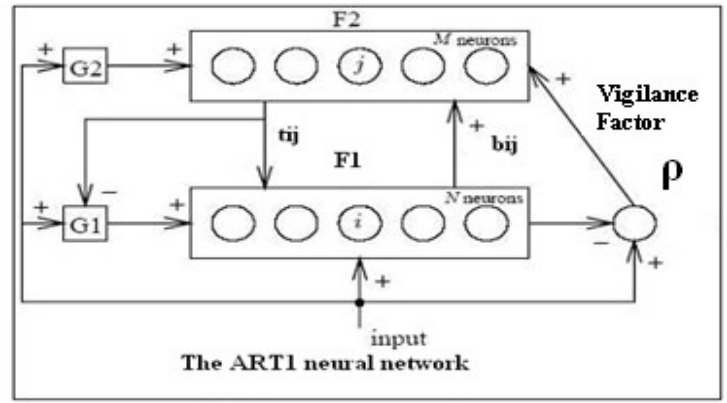

**Fig 3 The ART1 neural network**

For (ρ→0) the model is expected to have minimum number of clusters. As (ρ→1) it is expected to form maximum number of clusters. Some times to an extent of one input prototype vector in each cluster.

ART1 consists of comparison layer F1, recognition layer F2,and control gains G1 and G2. The input pattern is received at F1, whereas classification takes place in F2 Each neuron in the comparison layer receives three inputs: a component of the input pattern, a component of the feedback pattern, and a gain G1. A neuron outputs a 1 if and only if at least three of these inputs are high: the 'two-thirds rule.' Each input activates an output node at F2.The F2 layer reads out the top-down expectation to F1, where the winner is compared with the input vector. The vigilance parameter determines the mismatch that is to be tolerated when assigning each host a cluster. If the new input is not similar to the specimen, it becomes the second neuron in the recognition layer. This process is repeated for all inputs (parts) [9]. ART learns to cluster the input pattern by making the output neurons compete with each other for the right to react to a particular input pattern. The output neuron which has the weight vector that is most similar to the input vector claims this input pattern by producing an output of '1'and at the same time inhibits other output neurons by forcing them to produce '0's. in ART, only the winning node is permitted to alter its weight vector , which is modified in such a way that is brought near to the representative input pattern in cluster concerned.

ART neural networks are known for their ability to perform plastic yet stable on-line clustering of dynamic data set. ART adaptively and autonomously creates a new category. Another advantage of using ART1 algorithm to group users is that it adapts to the change in users' request access patterns overtime without losing information about their previous access pattern. The advantage comes from the fact that ART neural networks are tailored for systems with multiple inputs and multiple outputs, as well as for nonlinear systems. Besides this advantage, no extra detailed knowledge about the process is required to model using neural networks. The consideration of system dynamics leads to networks with internal and external feedback, where there are many structure variants.

```
ART1 Algorithm
Input:
    i.   Binary pattern vectors P_{c = 1 to m} each representing the
         multimedia VoD access patterns of the clients.
    ii.  Select a value for vigilance threshold between zero and one 0≤ρ≤ 1.
    iii. Initialize top-down weight t_{ij}(0) = 1 and bottom-up weight b_{ij}(0) = 1/(1+N)
         where N is the size of the input vector and also number of neurons in  F1.
Step 1 : Set up the network i.e., the input nodes.
Step 2 : Repeat steps 2 – 8 until all input vectors P_c are presented to the F1 layer.
Step 3 : Input the binary pattern X , X =(X_{i=1},X_{i=2},X_{i=3},X_{i=4}, ……..X_{i=N}) .
Step 4 : Calculate the "matching value" for every output node j .
                   N
         net_j = Σ Xi × b_ij   at the beginning j=1.
                  i=1
Step 5 : Determine the winning node j * the node in F2. net_j* = Max[net_j] .
                                                                 j
                                      Σ t_ij × Xi
Step 6 : Calculate "similar value" V_j* = ─────────
                                        Σ Xi
Step 7 : Conduct the vigilance test for the winning node
    Case 1 :  if V_j < ρ vigilance value
              This means that the input pattern is not similar to the connected
              weights and, hence it does not belong to this j* cluster.
              Find the next winning output node to see if it can pass the vigilance
              test. Otherwise, generate a new output node l in F2 layer.
              Initialize the top-down weights t_ij to the current input pattern
              Initialize bottom-up weights for the new node l.
                            Xi*
              b_ij(new) = ───────────
                          0.5+ Σ Xi*
                               i=1
    Case 2 :  if V_j ≥ ρ
              This means the input pattern matches to the output node j* node is the
              cluster for representing this pattern X. Update top-down weights
                                              t_ij * Xi
              t_ij* = t_ij* × Xi   and   b_ij = ─────────
                                              0.5 + Σ t_ij * .Xi
              end
Step 8 : Go to step 2.
Step 9 : End ART1_clustering() algorithm.
```

**Fig 4  An ART1 clustering Algorithm**

**4.5 Prefetching Scheme**

```
Procedure : ART1_based_prefetching (Client_Id Cid of the client requesting a video)
Preprocessing: Cluster the clients using the ART1_clustering algorithm. Each cluster is
represented as C_n where n is the number of clusters formed. The clusters C_1,C_2,C_3,C_4 ....C_n are
represented by prototype vectors. The prototype vectors for the kth cluster is of the form
Tk = (t_{k1},t_{k2},t_{k3},……t_{kn}) where t_{kj= 1…N} are the top-down weights corresponding to the node k in
layer F_2 of the ART1.
Input: Client_Id of the client requesting for a video.
Output: The array prefetched URLs[] which contains a list of URLs that are to be prefeched for
the client Cid.
Initialize count = 0
Step 1 : for n clusters formed using ART1_clustering algorithm
begin
Step 2 : Fetch the prototype vector of each cluster.
        for j= 1 to N do
           begin
Step 3 : if (t_{kj} = 1)       // j^{th} element of K^{th} vector.
           begin
               prefetched_URLs[count] = URLj
               count = count + 1
           end-if Step 3
        end-for Step 2
Step 4 : return prefetched _URLs[]
Step 5 : End_ART1_based_prefetching()
```

**Fig 5 ART1_based_Prefeching Algorithm**

Fig 4 above is the binary ART1 algorithm used to cluster the requests . The number of clusters formed depends on the vigilance factor. Fig 5 is ART1 based prefetching algorithm. At any time instance the videos are prefetched and stored in the cache prior to the request. The prototype selected will depend on the cluster formed at the same instance of time ,from the previous history. Those videos for which the binary value set to one in the prototype will be prefetched.

## 5  Results and Discussion

### 5.1 Performance of ART1 clustering technique

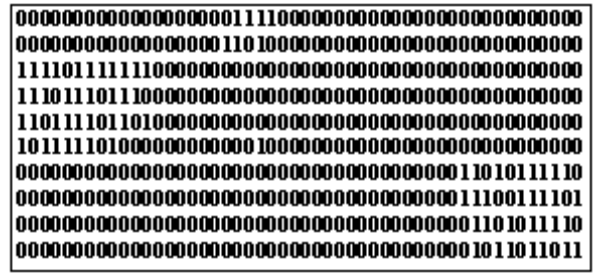

**Fig 6   A sample Input data**

Fig 6 above is a sample set of data from 10 different users.
A value 1 represents that more than two clients have requested for a video and a 0 no one has requested for that particular video.

A sample of 10 pattern vectors with N = 50 is taken from our simulation results and variation in number clusters formed with that of vigilance factor is observed[8]. The result is as shown in the graph below.  The number clusters formed increases with the increase in vigilance parameter. It
 clear from the graph that a value of 0.3 to 0.5 forms ideal number of clusters.

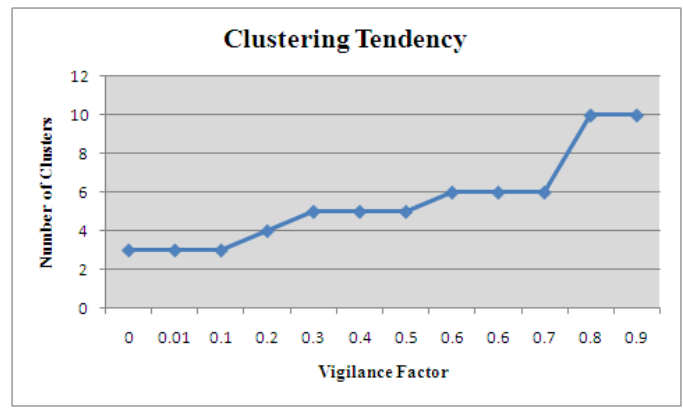

**Fig 7 Number of clusters verses vigilance parameter for sample input data**

The same when applied to 200 videos from different 50 clients with their frequency ranging  from  27  to  1530. The value of the vigilance parameter is   varied   between 0.30  to  0.50.

**Table 1 Number of Clusters formed by varying the value of vigilance   parameter**

| Vigilance Parameter | Number of Clusters |
|---|---|
| 0.3 | 18 |
| 0.35 | 23 |
| 0.40 | 30 |
| 0.45 | 34 |
| 0.475 | 39 |
| 0.50 | 42 |

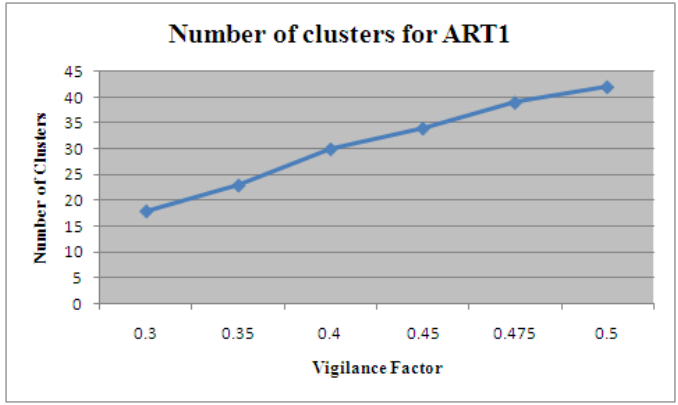

**Fig 8 Variation in the number of clusters formed by increasing the vigilance parameter**

### 5.2 Prefetching results

We have conducted an performance analysis based on two parameters. (a) Hits and (b) Accuracy. Hits indicate the number of videos that are requested from the prefetched videos, and accuracy is the ratio to hits to the number of videos being prefetched. Most of the techniques for Predictive Prefetching is for a single user. Whereas ART1 neural network based clustering reduces the overload on the network by prefetching the videos not for a single user, but for a community of users. It prefetches request with an accuracy as high as 92%. Each prototype vector of a cluster represents a possibility of community of users requesting for videos. This predictive prefetch is performed periodically over a sliding window defined by a session. An accurate predictions are difficult since customer behavior change over time [11]. The extremely popular videos are accessed frequently and the frequency may change not only on a daily or weekly, but even on an hourly basis. For eg children videos are likely to be popular early in the evening or in weekend morning [11], but less popular late at night.

The ART1 model reflects customer behavior over 24-hours period. From the historic data clusters are framed. Each cluster is represented by a prototype vector which is in the binary format and holds the information about most popular videos, requested by the members of that cluster. The historic data is collected at the end of each session. A session interval is defined on predefined parameter *maximum_idle_time*. Whenever a new request arrives, it is checked for the membership of matching cluster. All videos that are frequently requested by the members of that cluster might have been already present in the cache. Sometimes the new request may modify the prototype vector of a cluster. If a request arrives during a course of session the new prototype vector is computed using both historic data and also the current input vectors, so that accurate predictions can be done.

**Table 2 Result of prefetching scheme**

| Num of members each cluster | Videos prefetced | Hits | Accuracy % |
|---|---|---|---|
| 8 | 36 | 34 | 97% |
| 6 | 45 | 43 | 93% |
| 4 | 39 | 38 | 87% |
| 5 | 32 | 30 | 92% |

In Table 2, we have shown the result of our prefetching scheme by considering sample of four clusters. We prefetch the URLs for each client and verify the accuracy of our algorithm. The vigilance factor was set to 0.4. The average prediction accuracy of our scheme is 92.2%, which is considerably high.

### 6 Conclusion

In recent years, there have been a number of researches in exploring novel methods and techniques to group users based on the request access pattern. In this paper we have clustered and prefetched user request access pattern using ART1 clustering approach. The predictions were done over a time series domain. The proposed system has achieved good performance with high satisfaction and applicability. Effort on a way to improve by clustering the requests on demand.
The future work is on to count the number of requests that were considered to form any cluster. The prefetching algorithm will prefetch the most popular cluster at that instance which may improve the performance.

**P Jayarekha** holds M.Tech (VTU Belgaum ) in computer science securing second rank . She has one and a half decades experience in teaching field. She has published many papers. Currently she is working as a teaching faculty in the department of Information science and engineering at BMS College Of Engineering , Bangalore ,India.

**T.R. Gopalakrishnan Nair** holds M.Tech. (IISc, Bangalore) and Ph.D. degree in Computer Science. He has 3 decades experience in Computer Science and Engineering through research, industry and education. He has published several papers and holds patents in multi domains. He won the PARAM Award for technology innovation. Currently he is the Director of Research and Industry in Dayananda Sagar Institutions, Bangalore, India.